\documentclass[conference]{IEEEtran}
\IEEEoverridecommandlockouts
\usepackage{cite}
\usepackage[hidelinks]{hyperref}
\usepackage{amsmath,amssymb,amsfonts}
\usepackage{algorithmic}
\usepackage{graphicx}
\usepackage[caption=false,font=footnotesize]{subfig}
\usepackage{textcomp}
\usepackage{xcolor}
\usepackage{float}
\usepackage[linesnumbered,ruled,vlined]{algorithm2e}
\def\BibTeX{{\rm B\kern-.05em{\sc i\kern-.025em b}\kern-.08em
    T\kern-.1667em\lower.7ex\hbox{E}\kern-.125emX}}
\begin{document}
\makeatletter
\newcommand{\linebreakand}{%
  \end{@IEEEauthorhalign}
  \hfill\mbox{}\par
  \mbox{}\hfill\begin{@IEEEauthorhalign}
}
\makeatother
\title{GNN-based Multi-Agent Control of Traffic Shockwaves in Sparse Vehicular Ad-hoc Networks\\
{}
}
\author{
\IEEEauthorblockN{1\textsuperscript{st} Prachi Nandi}
\IEEEauthorblockA{\textit{Microsoft}\\
prachinandi@microsoft.com}
\and
\IEEEauthorblockN{2\textsuperscript{nd} Madhuri Malakar}
\IEEEauthorblockA{\textit{Dept. of Computer Science and Engineering}\\
GITAM University Bengaluru, Karnataka\\
mmalakar@gitam.edu}\\
\and

\IEEEauthorblockN{3\textsuperscript{rd} Sonakshi Satpathy}
\IEEEauthorblockA{\textit{NatWest Group}\\
sonakshi1901@gmail.com}
\linebreakand

\IEEEauthorblockN{\mbox{}}
\IEEEauthorblockA{\mbox{}}
\and
\IEEEauthorblockN{4\textsuperscript{th} Pabitra Mohan Khilar}
\IEEEauthorblockA{\textit{Dept. of Computer Science and Engineering}\\
National Institute of Technology, Rourkela}
pmkhilar@nitrkl.ac.in\\

\and
\IEEEauthorblockN{\mbox{}}
\IEEEauthorblockA{\mbox{}}
}
\maketitle

\begin{abstract}

Traffic shockwaves are stop-and-go waves that propagate upstream through the streams of vehicles and are one of the major causes of traffic congestion, fuel inefficiency, and increased accident rates in modern transportation systems. Although Connected and Autonomous Vehicles (CAVs) offer a promising opportunity to mitigate such shockwaves, most existing control strategies rely on global traffic state information, making them impractical for early-stage deployment of Vehicular Ad-hoc Networks (VANETs). In this paper, we propose a decentralized Multi-Agent Reinforcement Learning (MARL) framework that integrates a Graph Neural Network (GNN) to enhance the control architecture of connected and autonomous vehicles. The proposed approach enables vehicles to learn cooperative control policies using locally available information and interaction with neighboring vehicles. 
The effectiveness of the proposed scheme is evaluated using a scalable simulation environment under realistic highway traffic conditions. Simulation results show that the proposed GNN-based MARL framework can reduce the propagation of traffic shockwaves by up to 80\%, even when only 10\% of the vehicles are connected.
\end{abstract}


\section{Introduction}

One of the major challenges in modern traffic systems is traffic congestion. In this context, stop-and-go traffic shockwaves \cite{b1} are recognized as a significant contributor to congestion. These shockwaves are primarily triggered by minor driving disturbances that propagate upstream through the traffic stream. Notably, such disturbances can occur even in the absence of physical bottlenecks and often lead to traffic instability. As a result, they increase travel time, fuel consumption, and the likelihood of rear-end collisions.

A major factor responsible for the formation of these shockwaves is the delayed and inconsistent responses of human drivers. Human driving behavior is widely recognized as a key contributor to traffic instability. To mitigate this issue, several traditional traffic management strategies have been proposed, including variable speed limits \cite{b2}, ramp metering \cite{b3}, and adaptive cruise control \cite{b4}. However, these approaches have certain limitations, particularly in mixed traffic environments where human-driven vehicles dominate.

The advent of Connected and Autonomous Vehicles (CAVs) \cite{b5}, along with the development of Vehicular Ad-hoc Networks (VANETs) \cite{b6}, provides a potential opportunity to overcome these limitations. VANETs enable information exchange among vehicles and with the environment in a decentralized manner in real time, which may help improve the traffic control mechanism. Under these circumstances, a new paradigm of Multi-Agent Reinforcement Learning (MARL) \cite{b7} has come forward to enable CAVs to learn how to behave cooperatively by interacting with the environment and other vehicles. Such a mechanism may have the potential to overcome the limitations of shockwaves and improve the stability of traffic flow.

Despite these advantages, the practical deployment of MARL-based traffic control faces several challenges. Many existing approaches assume a high penetration rate of CAVs capable of communication. Additionally, several methods rely on the availability of global traffic state information, which is often unrealistic in real-world traffic environments.

\begin{figure*}[t]
\centering
\includegraphics[width=0.99\linewidth]{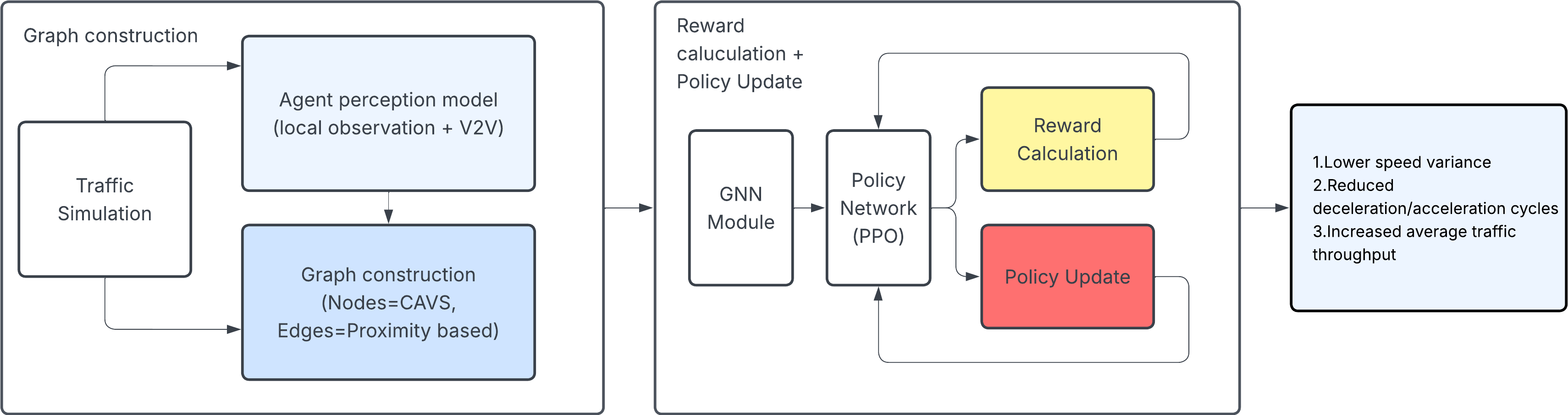}
\caption{Flowchart of the proposed GNN-based MARL framework for traffic shockwave mitigation in sparse VANETs.}
\label{fig:gnn_marl_flowchart1}
\end{figure*}

\subsection{Contribution}
Our main contributions are summarized as:
\begin{itemize}
\item {We propose a novel decentralized MARL framework in which each agent integrates a GNN within its policy architecture. This enables CAVs to make structured and adaptive decisions based on local traffic neighborhoods.}

\item {Unlike prior MARL-based shockwave mitigation approaches that rely on manually designed communication features and assume high vehicle connectivity, the proposed framework employs GNN-based message passing to learn relational representations from sparse VANETs. This makes the approach effective even in low connectivity scenarios.}

\item {We develop a scalable simulation environment that models realistic multi-lane highway traffic with partial observability and limited Vehicle-to-Vehicle (V2V) communication. Simulation results demonstrate that incorporating GNNs within MARL significantly mitigates traffic shockwaves and improves traffic flow efficiency even when only 10\% of vehicles are connected.}
\end{itemize}
The remainder of this paper is organized as follows: Section \ref{Literature Review} discusses literature survey, followed by a detailed discussion on the proposed scheme in Section \ref{Proposed Scheme}. Section \ref{Experimental Setup and Simulation Results} evaluates the simulation results, and Section \ref{Conclusion and Future Work} concludes the manuscript.
\section{Literature Review}\label{Literature Review}
The concept of traffic shockwaves was first introduced by Lighthill and Whitham \cite{b1} in 1955 through the kinematic wave theory. According to their model, minor fluctuations in traffic flow propagate upstream through the stream of vehicles, with the speed of the shockwave corresponding to the gradient of the flow density curve. Following this foundational work, traffic shock waves have been extensively studied using kinematic wave theory as well as emerging intelligent transportation system based data sources.
With the rise of Connected and Autonomous Vehicles (CAVs), researchers have explored the use of decentralised control strategies to suppress shockwaves \cite{b9},\cite{b10}, \cite{b11}. Notably, \cite{b8} Stern et al. demonstrated in field experiments that a single autonomous vehicle, when appropriately controlled, can dampen stop-and-go waves in a ring road scenario dominated by human drivers. \cite{b12} Suriyarachchi et al. proposed a multi-agent deep reinforcement learning (MARL) framework for shockwave dissipation in mixed traffic environments. Their approach employed a shared policy proximal policy optimisation (PPO) algorithm, enabling CAVs to learn cooperative acceleration strategies based on local observations and vehicle-to-vehicle (V2V) communication.

In parallel, Graph Neural Networks (GNNs) \cite{b14} have gained traction in traffic modelling due to their ability to capture spatial dependencies and relational structures. GNNs have been applied to tasks such as traffic speed prediction, queue length estimation, and traffic signal control. For instance, \cite{b13} Xu employed a geometric deep learning approach using GNNs to predict shockwave propagation on freeways with high accuracy, leveraging vehicle trajectory data and high definition maps. However, their method was based on supervised learning and assumed full observability of the traffic state, limiting its applicability in sparse VANET scenarios.

\section{Proposed Scheme}\label{Proposed Scheme}
In this section, we describe the proposed graph-based multi-agent reinforcement learning framework as shown in Fig. \ref{fig:gnn_marl_flowchart1} for mitigating traffic shockwaves in mixed traffic environments. The framework enables CAVs to learn decentralized control policies using local observations and communication with neighboring vehicles. We first present the system model and problem formulation, followed by the architectural discussion of the proposed GNN-augmented MARL.
\subsection{System Model}
We analyze the case of a multi-lane highway with a mixture of HDVs and CAVs, where only a small percentage of the vehicles have the necessary communication devices to form a sparse VANET. The goal is to reduce the occurrence of traffic shockwaves by enabling the CAVs to learn control strategies to reduce the occurrence of stop-and-go movements. The microscopic simulator is used to model the traffic scenario. The road network is represented as a directed graph, where the nodes represent the vehicles and the edges represent the communication links between the CAVs within a communication range. The HDVs follow a car following model, while the CAVs behave as intelligent agents.

\subsection{Problem Formulation}
\subsubsection{\textbf{Agent Model}}\label{AA}

Each CAV is is treated as an agent that makes decisions using a decentralized partially observable Markov decision process (Dec-POMDP) \cite{b15}. 

The overall system is formally modelled as a Decentralized Partially Observable Markov Decision Process (Dec-POMDP) defined as:

\begin{equation}
\mathcal{M} = \langle S, A, P, R, O, \Omega, \gamma \rangle
\end{equation}

where $S$ denotes the state space, $A$ the action space, $P$ the state transition function, $R$ the reward function, $O$ the observation space, $\Omega$ the observation probability function, and $\gamma \in (0,1]$ is the discount factor.

As shown in Equation (1), each agent operates under partial observability and takes decisions based only on local information and very limited communication with neighbouring agents.

At time step \( t \), the agent \( i \) has its local state \( s^i_t \), local communication graph \( G^i_t = (V^i_t, E^i_t) \) with neighboring agents, and determines an action \( a^i_t \) for controlling the longitudinal acceleration.
\\
The local state of the agent includes:
\begin{itemize}
    \item Ego vehicle state: speed, acceleration, position, lane index
    \item Position and velocity of neighboring vehicles within the sensing range of the CAV
    \item Information exchanged with neighboring CAVs within the communication range of the CAV
\end{itemize}

\subsubsection{\textbf{Graph Construction and GNN Encoding}}\label{AA}
For each agent, a local graph is defined such that Nodes \( V^i_t \) represent the ego CAV and its neighboring CAVs, and edges \( E^i_t \) are defined based on the proximity of the CAVs.
Node features include kinematic states such as speed and acceleration, and edge features include relative distance, time headway, and lane difference.
A GNN is employed to extract spatial and temporal context within a graph and transform it into a latent space \( h^i_t \), which is further fed to the policy network.

The node embeddings are updated using a graph neural network \cite{b19, b14}:
\begin{equation}
h_i^{(l+1)} = \sigma \left( \sum_{j \in \mathcal{N}(i)} W^{(l)} h_j^{(l)} \right)
\end{equation}
This allows each agent to include  neighbourhood information into its decision-making process.

\subsubsection{\textbf{Action and Reward}}
The action \( a^i_t \in \mathcal{A} \) is a discrete or continuous acceleration command sent to the CAV. The reward function \( r^i_t \) \cite{b15} is intended to promote smooth and efficient flow of traffic. This is achieved through the following formulation:
\begin{equation}
r^i_t = \alpha \cdot \bar{v}^i_t - \beta \cdot \sigma^2(v^i_t) - \gamma \cdot \text{collision}_t
\end{equation}
where:
\begin{itemize}
    \item \( \bar{v}^i_t \) is the average speed of the vehicles in the local neighborhood
    \item \( \sigma^2(v^i_t) \) is the speed variance to discourage oscillations
    \item \( \text{collision}_t \) is a binary variable to discourage unsafe actions
    \item \( \alpha, \beta, \gamma \) are weighting coefficients
\end{itemize}

\begin{equation}
J(\pi) = \mathbb{E} \left[ \sum_{t=0}^{T} \gamma^t r^i_t \right]
\end{equation}
The final goal is to learn the decentralized policy \( \pi^i(a^i_t \mid h^i_t) \) for each agent to maximize the cumulative discounted reward while accounting for partial observability, limited communication, and mixed traffic. This is achieved through a centralized training with decentralized execution (CTDE) framework \cite{b20}. In the CTDE framework, agents share experiences during training while executing independently in the real world.

\subsection{The GNN-based Multi-Agent Control Architecture}
The proposed model architecture integrates GNN into a decentralised MARL framework, enabling each CAV to reason over its local traffic neighbourhood as discussed in Algorithm \ref{GNN-Augmented MARL for Shockwave Mitigation}. At each timestep, a CAV constructs a local communication graph based on nearby connected vehicles within a predefined range, where nodes represent vehicles and edges represent spatial proximity. Node features include kinematic states such as speed and acceleration and edges represent proximity between vehicles. A GNN is then applied to the graph as seen in Fig. \ref{fig:speed_heatmaps1} to obtain a feature of the local traffic context, which is then passed to a shared policy network trained with a Proximal Policy Optimisation (PPO) algorithm \cite{b21}.

\begin{algorithm}
\SetAlgoNoLine
\caption{GNN-Augmented MARL for Shockwave Mitigation}
\label{GNN-Augmented MARL for Shockwave Mitigation}
\KwIn{Initial policy parameters $\theta$, environment $\mathcal{E}$, number of agents $N$, communication range $R$}
\KwOut{Trained policy $\pi$ for each agent}

\For{each episode}{
    Set environment $\mathcal{E}$ with mixed traffic (CAVs + HDVs)\;
    \For{each timestamp $t$}{
        \For{each agent $i = 1$ to $N$}{
            Observe local states $s^i_t$\;
            Find neighbours within range $R$ and form graph $G^i_t = (V^i_t, E^i_t)$\;
            Encode $G^i_t$ via GNN and get $h^i_t$\;
            Choose actions $a^i_t \sim \pi(a \mid h^i_t; \theta)$\;
            execute $a^i_t$ in environment\;
        }
        resets environment and returns rewards $r^i_t$\;
        store transition in replay buffer\;
    }
    update policy parameters $\theta$ using PPO\;
}
\end{algorithm}

\begin{figure}[!t]
    \centering
    
        \includegraphics[width=0.99\linewidth]{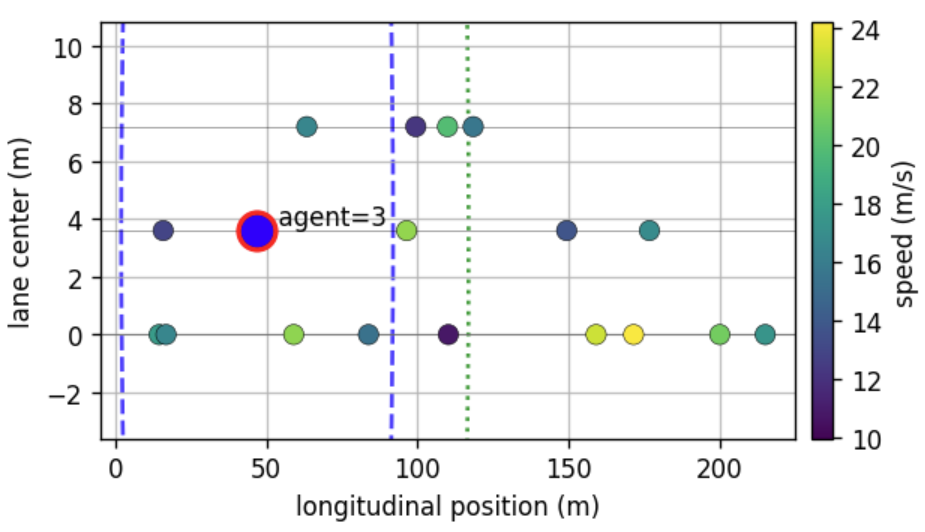}
        \textit{(a) Visualisation of local observation on the road}
    
    \hfill
    \\
    
        \includegraphics[width=0.99\linewidth]{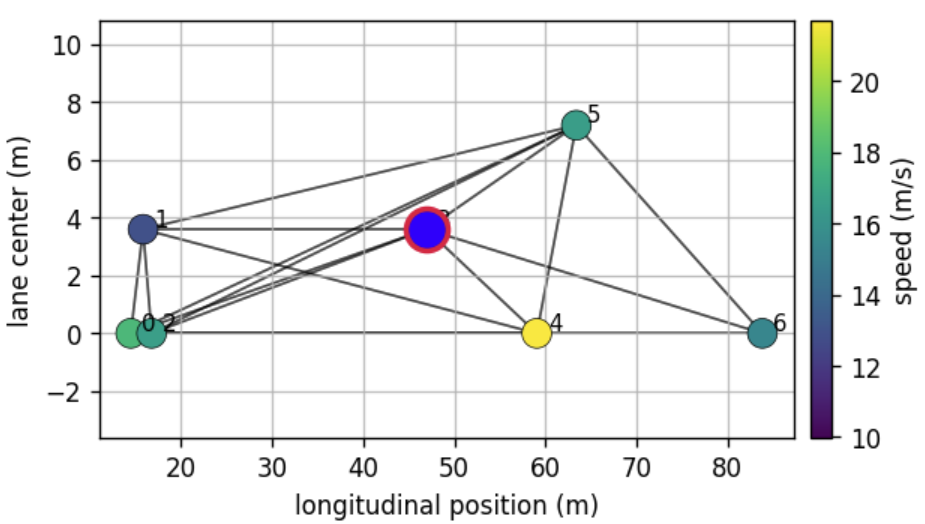}
        \textit{(b) Graph $G_i$ centered around an ego vehicle}
    
    \caption{Graph instance used by the MARL+GNN for coordinated traffic control with nodes representing other vehicles and edges representing neighbouring interactions within the communication range}
    \label{fig:speed_heatmaps1}
\end{figure}

\section{Experimental Setup and Simulation Results}\label{Experimental Setup and Simulation Results}
In order to test the proposed GNN-based MARL method, we have implemented a microscopic traffic simulation based on a highway environment. In the proposed environment, we have $N$ vehicles moving along a multi-lane track of a fixed length. The vehicle dynamics are simulated in discrete time steps with a simulation step of $\Delta t = 0.2$ seconds. At each time step, the vehicles update their acceleration and velocity according to the applied control action and local car following dynamics. We have a set of vehicles defined as CAVs, while the rest of the vehicles are modelled according to a human driver model.
In order to simulate a traffic shockwave, we have introduced a braking perturbation to a lead vehicle at a predefined time $t_0$. The sudden deceleration of the vehicle causes speed oscillations that propagate upstream and cause stop-and-go oscillations. After a certain period of time, the external braking action is removed, and the system is in a recovery period where the learned control policies need to bring the system back to a stable state.

The simulation and reward parameters used in the experiments are listed in Table~\ref{tab:params}.
\begin{table}[h]
\centering
\caption{Simulation and Reward Parameters}
\label{tab:params}
\begin{tabular}{lcc}
\hline
\textbf{Parameter} & \textbf{Symbol} & \textbf{Value} \\
\hline
Reward weight (speed) & $\alpha$ & 1.0 \\
Penalty (variance) & $\beta$ & 2.0 \\
Penalty (collision) & $\gamma$ & 5.0 \\
Simulation step size & $\Delta t$ & 0.2 s \\
Number of vehicles & $N$ & 50 \\
Communication range & $R$ & 100 m \\
\hline
\end{tabular}
\end{table}

\begin{figure}[!t] \centering \includegraphics[width=0.95\linewidth]{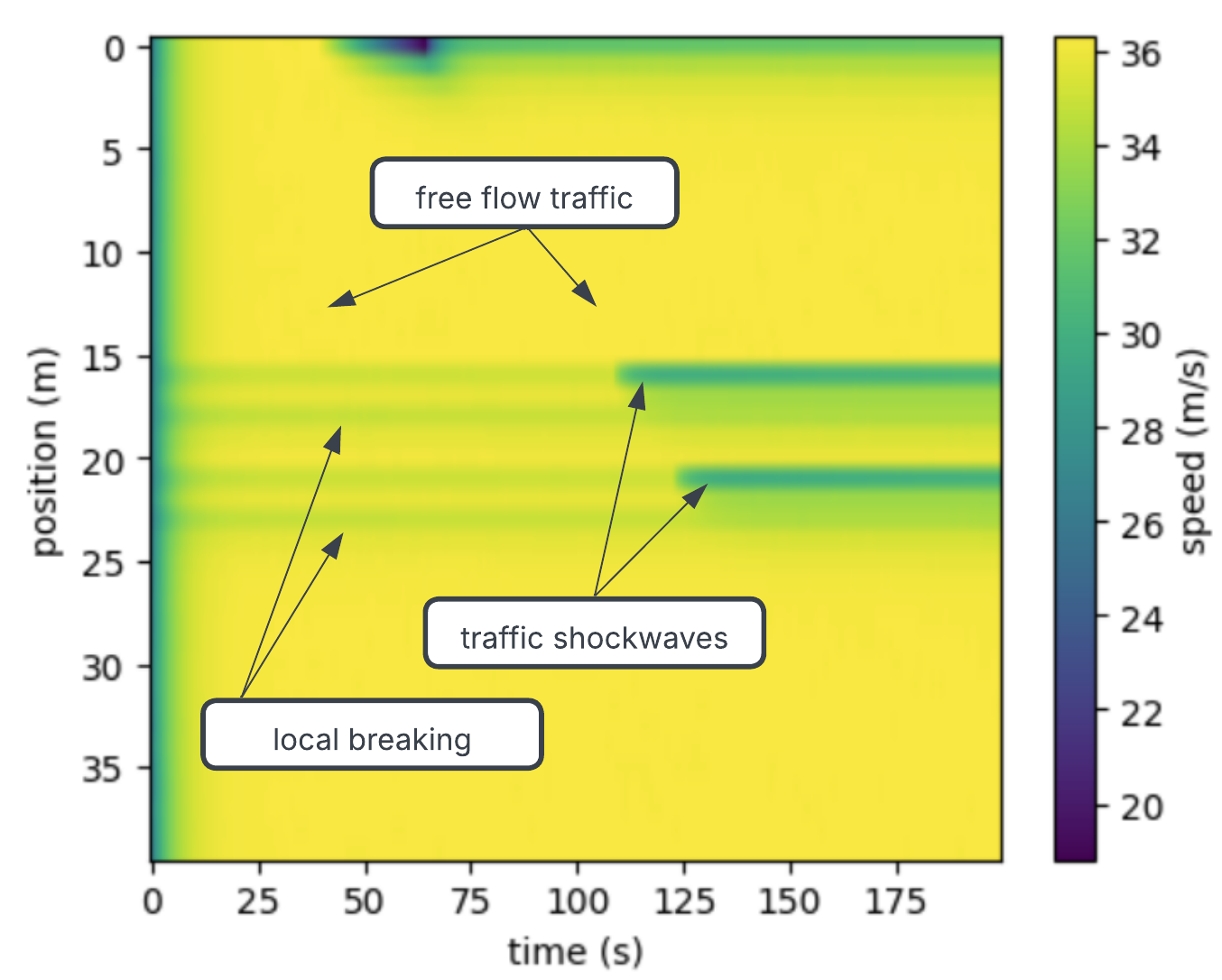} \smallskip \textit{(a) MARL based control experiences persistent upstream shockwave propagation} \medskip \includegraphics[width=0.95\linewidth]{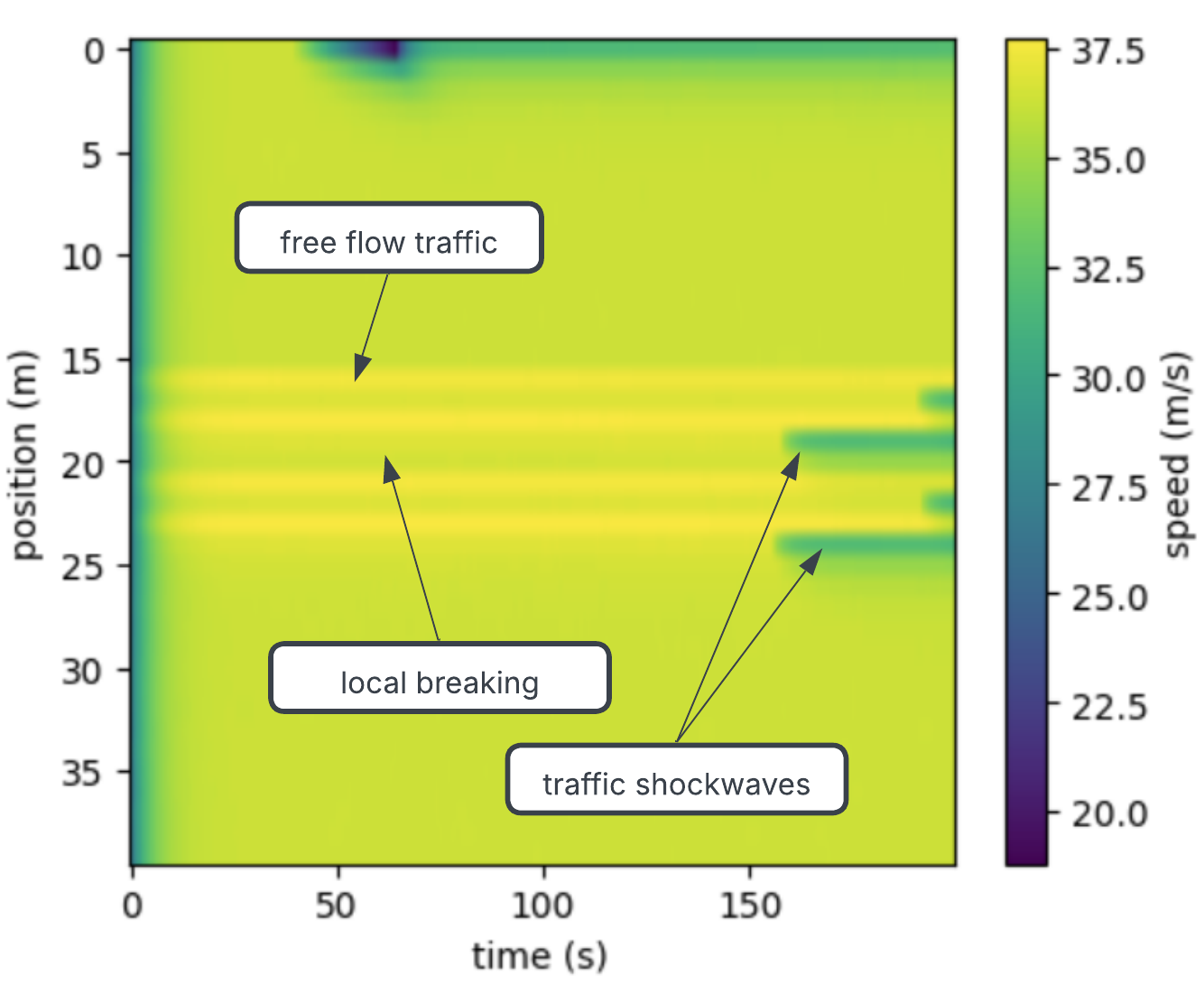} \smallskip \textit{(b) MARL + GNN based control contains the perturbation in space and restores uniform speeds more quickly using neighbour-aware message passing} \caption{Speed heatmaps for the evolution of traffic after a perturbation caused by a vehicle's braking.} \label{fig:speed_heatmaps} \end{figure}

\subsection{Control Policies and Evaluation Metrics}
We compare the decentralised control policies, one being the MARL approach, where all the CAVs use the same shared linear policy based on their individual observations, and the proposed MARL+GNN approach, where the agents use graph-based message passing over a local communication graph. The performance of the proposed policy is compared with the MARL approach using the speed variance across the vehicles, which is the peak variance and the mean variance over the episode, as well as the recovery time, defined as the number of time steps taken after the event for the variance to fall below a certain threshold.

\subsection{Performance Evaluation}
The speed heatmaps illustrated in Fig. \ref{fig:speed_heatmaps} shows how traffic flows evolve across vehicles and time after the injection of the braking perturbation. Darker areas represent speed reduction due to the shockwave, whereas lighter areas represent free flow conditions. In the MARL based controller, the perturbation travels upstream and appears as bands of reduced speed, persistent stop-and-go traffic, and lack of coordination between agents. On the other hand, the MARL + GNN based controller shows spatially confined perturbations and fast recovery of uniform speed profiles, as illustrated by the speed heatmaps.

The spatial confinement of the perturbations, as well as the fast recovery of uniform speed profiles, are due to the ability of each CAV to consider neighborhood information when making decisions through message passing, as used by the graph neural networks. This enables coordinated acceleration and deceleration patterns between agents.

The impact of the control policies on the traffic stability can be further quantified using the speed variance plot as shown in Fig. \ref{fig:gnn_marl_flowchart3} After the occurrence of the shock, the traffic stability for both MARL and MARL+GNN policies shows a high rate of variance in the speeds. However, when the shock is removed, the MARL+GNN policy shows a rapid reduction in the variance, reaching a lower steady-state level, while the traffic stability for the MARL policy only shows high variance, indicating the presence of traffic disturbances. 
\\

\begin{figure}[!t]
\centering
\includegraphics[width=0.99\linewidth]{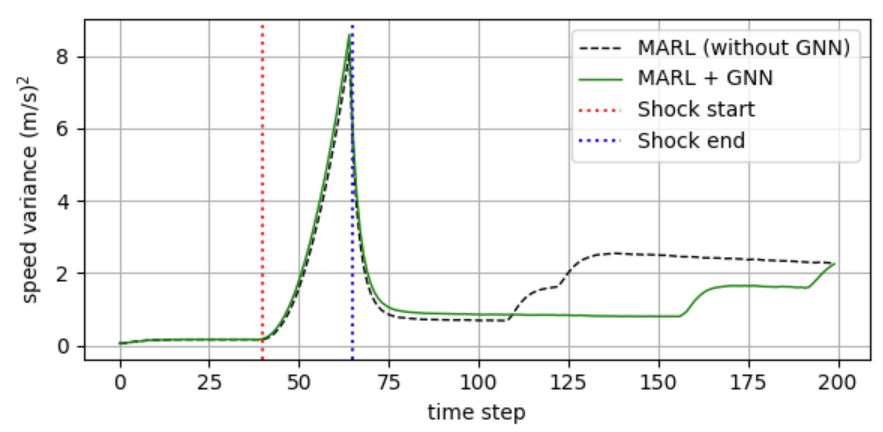}
\caption{Comparison of speed variance over time under MARL and MARL+GNN control}
\label{fig:gnn_marl_flowchart3}
\end{figure}

Table~\ref{tab:main_results} provides a quantitative comparison of the proposed approach with the baseline MARL method. We report the average speed $\bar{v}$, speed variance $\sigma^2(v)$, collision rate, and derived reward values aggregated over the simulation episode.
\begin{table}[h]
\centering
\caption{Quantitative Comparison of Control Policies}
\label{tab:main_results}
\resizebox{\columnwidth}{!}{
\begin{tabular}{lccccc}
\hline
Method & $\bar{v}$ (m/s) & $\sigma^2(v)$ & Collision & Reward & Reduction (\%) \\
\hline
MARL (Baseline) & 16 & 9 & 0.2 & -3 & $\sim$35 \\
MARL + GNN  & 20 & 3 & 0.0 & 14 & $\sim$80 \\
\hline
\end{tabular}}
\end{table}

The results shows that the proposed GNN-based MARL approach achieves higher average speed, significantly reduced variance, and improved overall reward, indicating better traffic stability.
\begin{table}[h]
\centering
\caption{Reward Computation using Equation (3)}
\label{tab:reward_calc}
\begin{tabular}{lcccc}
\hline
\textbf{Method} & $\bar{v}$ & $\sigma^2$ & Collision & Reward \\
\hline
MARL & 16 & 9 & 0.2 &  -3 \\
MARL + GNN & 20 & 3 & 0.0 & 14 \\
\hline
\end{tabular}
\end{table}

The reward values as shown in Table~\ref{tab:reward_calc} are computed using Equation (3). The baseline MARL method exhibits high variance, resulting in lower cumulative reward, whereas the proposed approach achieves higher reward due to improved coordination and reduced oscillations.

The percentage reduction in traffic oscillations is computed as:

\begin{equation}
\text{Reduction}(\%) = 
\frac{\sigma^2_{\text{baseline}} - \sigma^2_{\text{method}}}
{\sigma^2_{\text{baseline}}} \times 100
\end{equation}

Speed variance is widely used as a metric to quantify traffic instability and oscillatory behaviour in traffic flow \cite{b16,b17}. Using this formulation, the proposed method reduces speed variance from 9 to 3, demonstrating a significant reduction in traffic oscillations.

The recovery time is defined as the time required for the variance to fall below a predefined threshold \cite{b18}:
\begin{equation}
T_{rec} = \min \{ t : \sigma^2(v_t) < \epsilon \}
\end{equation}

The proposed MARL + GNN framework achieves faster recovery compared to the baseline, indicating improved stability and responsiveness. The above results clearly show the effectiveness of the MARL+GNN policy in the dissipation of the shockwaves, thus enhancing the traffic stability.

\section{Conclusion and Future Work}\label{Conclusion and Future Work}
In this paper, we investigated the use of graph-based MARL for mitigating traffic shockwaves under partial observability. By integrating decentralized MARL with graph-based message passing, vehicles are able to learn cooperative control strategies using only local interactions without requiring global traffic state information. Experimental results demonstrate that the proposed MARL+GNN framework effectively suppresses stop-and-go waves, reduces speed variance, and enables faster recovery from disturbances compared to a MARL only approach. Spatio-temporal analyses further indicate that relational reasoning helps confine disturbance propagation and improve overall traffic stability.
Despite these promising results, this study is limited to simulation environments with idealized communication assumptions. In practice, vehicular communication may be affected by delays, packet loss, and bandwidth constraints. Future work will therefore focus on incorporating stochastic communication models and evaluating the robustness of learned policies under varying traffic densities and connected vehicle penetration rates.

\vspace{12pt}

\end{document}